\documentclass[
]{ceurart}

\pdfoutput=1
\usepackage{listings}
\usepackage{xcolor}
\usepackage{hyperref}
\usepackage[justification=centering]{caption}

\begin{document}


\copyrightyear{2024}

\copyrightclause{Copyright for this paper by its authors. Use permitted under Creative Commons License Attribution 4.0 International (CC BY 4.0).}

\conference{Edge AI meets Swarm Intelligence Technical Workshop, September 18, 2024, Dubrovnik, Croatia}

\title{Towards an Implementation of the Knowledge-Based Control Plane for Intelligent Swarm Networks}


\author{Xuanchi Guo}[
orcid=0009-0003-7448-022X,
email=x.guo@tu-berlin.de
]

\author{Anh Le-Tuan}[
orcid=0000-0003-2458-607X,
email=anh.letuan@tu-berlin.de,
]

\author{Danh Le-Phuoc}[%
orcid=0000-0003-2480-9261,
email=danh.lephuoc@tu-berlin.de,
]

\address{Technical University of Berlin, Germany}


\begin{abstract}
    This paper proposes the possibility of integrating Dynamic Knowledge Graph (DKG) with Software-Defined Networking (SDN). This new approach aims to assist the management and control capabilities of the swarm network. The DKG works as a unified network data view, capturing network information such as topology, flow rules, host information, switch information, link status, and in-band network telemetry (INT) data. Benefited from the deep programmability of SDN, the network information can be converted into RDF format constantly, and the DKG will be dynamically updated. This approach helps the network operators to control their network infrastructure, such as allocating resource effectively and decision making at the application layer. Potential use cases demonstrate the applicability and advantages of the proposed approach. Examples include access control in swarm network scenarios and applying adaptive routing strategies, etc. These use cases illustrate how DKG-based SDN can address swarm network management challenges effectively, optimizing performance and resource utilization.
\end{abstract}


\maketitle

\section{Introduction}

The \textit{knowledge plane (KP)}~\cite{clark2003knowledge} or the \textit{knowledge-defined networking (KDN)} refers to the paradigm of integrating knowledge-based systems, such as knowledge graph (KG), systems with \textit{software-defined networking (SDN)}~\cite{kreutz2014software} to enhance the automation and operability of network elements.  


Typically, SDN vertically divides the network into three planes of functionality: the data plane, the control plane, and the management plane. The data plane consists of network devices (switches/routers) responsible for storing, forwarding, and processing data packets. The control plane represents the protocols used for defining the matching and processing rules of the data plane elements. The management plane, or application plane, includes software used for monitoring and configuring the control functionality. The network policy is defined in the management plane, the control plane enforces the policy, and the data plane executes it by forwarding data accordingly. KDN proposes adding a knowledge plane between the control plane and the management plane to facilitate understanding the network's behavior and automatically operating the network accordingly.


P4~\cite{bosshart2014p4} is a language designed for programming the data plane of network devices. It allows for introducing customized packet processing rules. In the context of swarm networks, it is supposed that all switches and routers function as P4 switches, which enables dynamic changes in packet forwarding rules and the implementation of new network protocols as needed. The P4Runtime API~\cite{P4RuntimeSpecs} is a control plane specification, providing standarization for controlling the data plane elements which are defined in a P4 program. P4Runtime facilitates the communication between the control plane and the P4 based data plane.

In swarm networks, we particularly focus on the coordination of networks of many simple, autonomous devices like robots and vehicles. This paradigm of swarm networks is closely related to swarm intelligence, as network operators require comprehensive knowledge of the swarm networks to make intelligent decisions. By leveraging the new paradigm of KDN, this paper aims to utilize and explore the capabilities of DKG to adapt to changing network conditions, optimize performance, and enhance the coordination and intelligence for swarm networks. The adoption of DKG not only improves the operational efficiency of the swarm network but also ensures robust and scalable management of these networks.

This paper is structured as follows. First, in section \ref{related works}, it summarizes related work on the concept of KG in SDN. Then, it provides an overview of the DKG in SDN, detailing its role, the key APIs used, and the design architecture of the DKG, in section \ref{overview}. Next, section \ref{use cases} presents use cases to demonstrate the feasibility of DKG for communication in swarm networks. Finally, the paper concludes with a summary in section \ref{conclusion}.

\section{Related Work} \label{related works}

Inspired by the vision of Knowledge plane~\cite{clark2003knowledge}, numerous studies have been done to introduce the abstraction of Knowledge graph into the network management tasks, especially with the rise of artificial intelligence. Most of these works emphasize theoretical and architectural designs rather than practical experiments on real-world network testbeds~\cite{xiao2006integration}~\cite{kanelopoulos2010ontology}~\cite{ohshima2015construction}~\cite{quinn2016knownet}~\cite{houidi2016knowledge}. For example, KnowNet~\cite{quinn2016knownet} employs Knowledge Graphs to capture data and information about the network, creating network management applications capable of managing and reasoning over this knowledge. However, KnowNet primarily targets enterprise network management and lacks detailed implementation specifics, such as the knowledge construction process.

On the contrary, SeaNet~\cite{zhou2021seanet}, is a knowledge graph-driven method for autonomic management of SDN. It provides a practical implementation of their proposed methodology and includes experiments to evaluate their solution with various network management use cases. The main components of SeaNet contains a knowledge base generator, a SPARQL engine\footnote{\url{https://www.w3.org/TR/sparql11-query}}, and a network management API. A distinguishing feature of SeaNet is its use of TOCO \footnote{\url{https://github.com/QianruZhou333/toco_ontology.git}}, an ontology for telecommunication networks that describes resources in heterogeneous telecommunication environments, as the language to construct their knowledge graph. To the best of our knowledge, this proposal is the first to deploy a knowledge graph to assist the ONOS controller in managing the P4 switch data plane within a swarm network scenario.

\section{Overview of DKGs in SDN} \label{overview}

This section provides an overview of the DKG in SDN. It discusses the role of the DKG within an SDN controller in section \ref{dkg in onos}, the P4Runtime API which enables interaction between the SDN controller and controlled devices in section \ref{p4 runtime}, and the architecture of the proposed DKG in section \ref{architecture}.

\subsection{The Role of DKGs in an ONOS Controller} \label{dkg in onos}

\begin{figure}[ht!]
    \centering
    \includegraphics[width=0.65\linewidth]{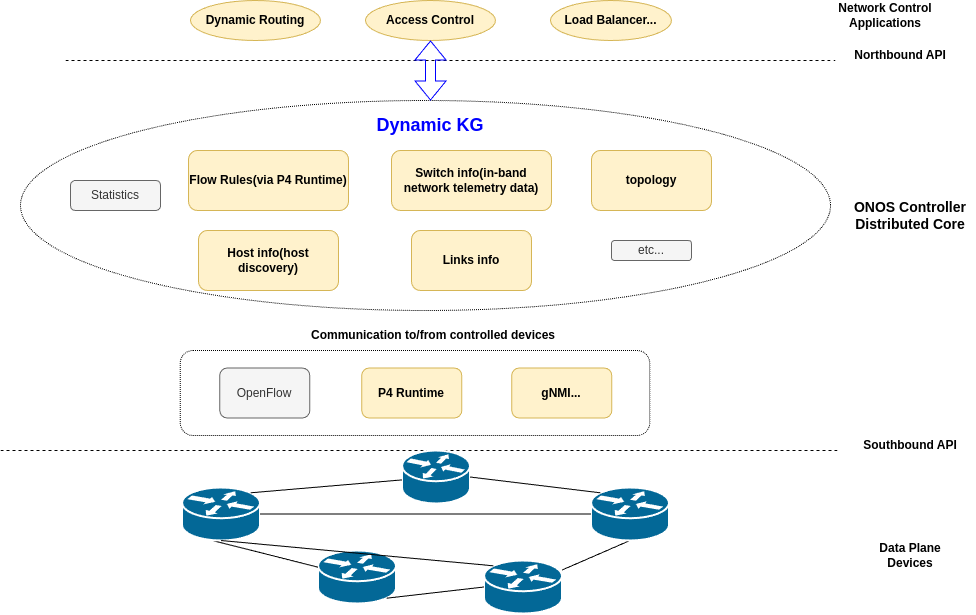}
    \caption{DKG in a simplified view of ONOS Controller}
    \label{fig:dkg in sdn}
\end{figure}

To begin, we examine the specific role that the DKG fulfils within the ONOS controller. Figure~\ref{fig:dkg in sdn} illustrates the DKG within a traditional SDN controller. According to Kurose et al.~\cite{KuroseRoss16}, the functionality of a canonical SDN controller can be divided into three layers: the communication layer, the network-wide state management layer, and the interface to the network control application layer, arranged from bottom to top.

SDN controllers are implemented in various languages, with OpenDaylight\footnote{\url{https://docs.opendaylight.org/en/stable-potassium/}} and ONOS\footnote{\url{https://opennetworking.org/onos/}} being particularly notable for their open-source nature and extensive industrial adoption. For this paper, we will use the ONOS controller to represent a typical SDN controller. ONOS, which stands for Open Network Operating System, serves as the control plane for a software-defined network. It operates as a logically centralized remote controller, capable of running as a distributed system across multiple servers. ONOS provides APIs that simplify the development of applications to manage network operations, offering benefits such as scalability, high availability, and robust performance.

In Figure~\ref{fig:dkg in sdn} which depicts a simplified view of ONOS controller, the layers are detailed as follows:

\begin{itemize}

    \item Southbound API: This layer manages communication between the ONOS controller and the network devices within the data plane. It utilizes a set of southbound protocols, including P4 Runtime, gNMI\footnote{\url{https://github.com/openconfig/gnmi}}, and gNOI\footnote{\url{https://github.com/openconfig/gnoi}}, to facilitate interactions with underlying hosts, links, and switches.
    
    \item ONOS Distributed Core: This layer maintains the state of the network's links, hosts, and switches. ONOS enhances scalability and availability by deploying across multiple servers, each containing an identical copy of the software. Through coordination and replication mechanisms, ONOS ensures a scalable and highly available system, providing a logically centralized view of core services to both upper-layer applications and data plane devices. The DKG extends the ONOS distributed core by storing network state information, such as topology, hosts, and devices. DKG's APIs allow various types of information to be converted into DKG-compliant formats for storage in its knowledge graph. The core of DKG's design is the seamless integration with the distributed core.
    
    \item Northbound API: These interfaces enable the controller to interact seamlessly with network control applications, allowing these applications to gain insights from the network state management layer. This capability enables them to make informed decisions and take appropriate actions. The DKG is designed to enhance this interaction by providing an alternative abstraction of network state information. With the DKG, network control applications can perform dynamic routing based on topology information, implement access control measures, and achieve load balancing. This system allows applications to query relevant knowledge and update their states as necessary based on the information retrieved from the knowledge base, ensuring efficient and adaptive network management.
\end{itemize}
    
\subsection{P4 Runtime API: Facilitating ONOS Controller Interaction with P4 Switches}\label{p4 runtime}

\begin{figure}[ht!]
    \centering
    \includegraphics[width=0.6\textwidth]{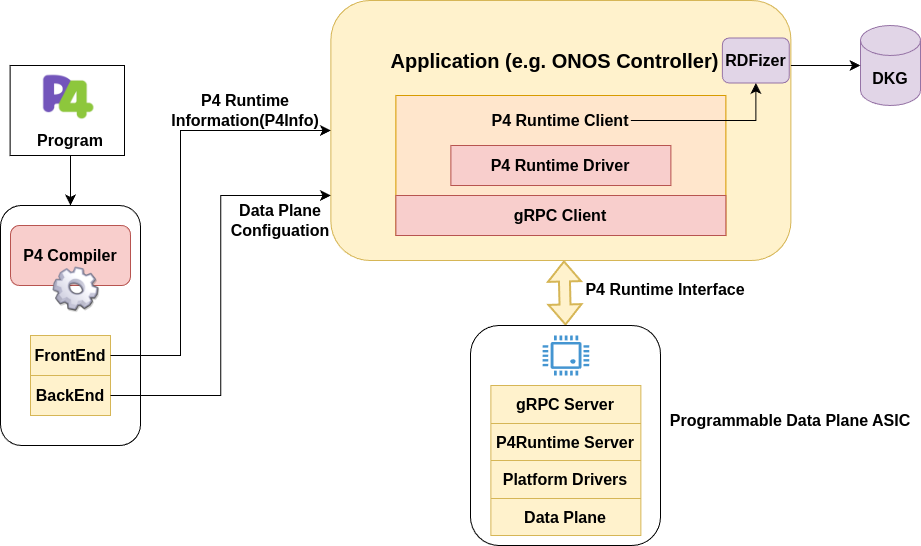}
    \caption{RDFizer in a Single Remote P4 Controller of a P4 Target}
    \label{fig:p4_controller_target}
\end{figure}

The P4 Runtime API \footnote{\url{https://p4.org/p4-spec/p4runtime/main/P4Runtime-Spec.html}} is crucial for facilitating communication between the ONOS controller and the underlying P4 switches. This protocol establishes seamless communication channels, with controllers acting as gRPC clients and data plane elements operating as gRPC servers. According to the P4 Runtime Specification~\cite{P4RuntimeSpecs}, "P4Runtime API is a control plane specification for controlling the data plane elements of a device defined or described by a P4 program." Typically, P4Runtime supports multiple controllers, while in Figure~\ref{fig:p4_controller_target}, we only illustrate a single remote controller managing a P4 target, with one ONOS controller interacting with the P4 target via the P4Runtime protocol. 

Within this framework, the RDFizer plays a crucial role, as depicted in Figure~\ref{fig:p4_controller_target}. Its primary function is to extract essential metadata such as packet header information from original packets sent to the controller for further processing. Additionally, the RDFizer can retrieve P4Runtime control entities, such as table entries from the controlled P4 switch, which are accessible via the P4Runtime API. The RDFizer then encapsulates this data into RDF objects, which are dynamically stored in knowledge graphs during runtime.

ONOS has been extended to allow users to integrate their custom P4 programs. It also supports new applications to control custom or new protocols defined in the P4 program. Typically, a P4 program is compiled for a specific P4 target into a JSON format file and loaded onto the respective P4 target. In this setup, the P4 switch operates as a gRPC/P4Runtime server, while the controller acts as the gRPC/P4Runtime client, capable of modifying table entries through P4Runtime APIs. Key switch-to-controller messages in P4Runtime include:
\begin{itemize}
    \item Modifying table entries, containing adding, updating, reading, deleting table entries;
    \item Packet in/packet out services, denoting the controller can send packets towards a specific switch port to the P4 switch or the P4 switch transferring packets to the controller.
\end{itemize}

These key control entities in P4Runtime are also supported by the ONOS northbound API. For instance, table entries in P4Runtime correspond to ONOS's \emph{Flow Rule Service} and \emph{Flow Objective Service}, while P4Runtime's Packet-in/out services are equivalent to ONOS's \emph{Packet Service}.

Additionally, the integration of packet I/O mechanisms enables the streaming of packets from the data plane to the control plane via dedicated CPU ports. This allows for further inspection and analysis of network traffic, providing administrators with deeper insights into network performance and security. A small P4 code snippet illustrating the packet I/O services enabled in a P4 program is shown in Appendix Figure~\ref{fig:p4 code}, with header declarations of types \emph{cpu\_in} and \emph{cpu\_out} presented in Appendix Figure~\ref{fig:p4 header}.

The central part of the process of RDFizing is the conversion of network metadata associated with P4 into structures that adhere to the RDF standards. The use of dynamic knowledge graphs allows for complex queries and analyses on the network's behavior and characteristics to be performed effectively. Additionally, because the information in these knowledge graphs is both accessible and updatable, it can adapt to the constantly evolving nature of swarm network environments.

\subsection{Architecture of the Proposed DKG Based SDN}\label{architecture}

\begin{figure}[ht!]
    \centering
    \includegraphics[width=0.6\textwidth]{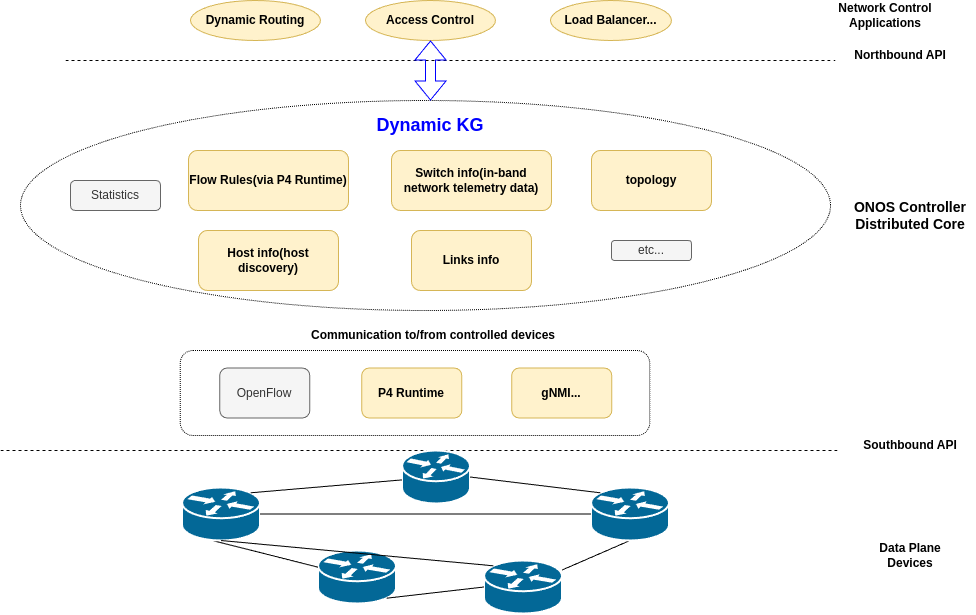}
    \caption{Architecture Overview of Dynamic Knowledge Graph}
    \label{fig:dkg-arch}
\end{figure}

This section will focus on the architecture of the proposed DKG and its integration with other components such as the ONOS controller and controlled P4 devices in the data plane, as illustrated in Figure~\ref{fig:dkg-arch}. The key components depicted include the knowledge generator, referred to as the RDFizer, the dynamic knowledge graph, and its corresponding SPARQL engine.

The ONOS controller interacts with real-time data plane sources to retrieve data, which can originate from packets or P4Runtime control entities, and abstracts this data into Plain Old Java Objects (POJOs). The RDFizer acts as a knowledge generator by transforming these POJOs into a semantically uniform format, specifically the Resource Description Framework (RDF)\footnote{\url{https://www.w3.org/RDF/}}. Essentially, the RDFizer structures the data into a consistent RDF format that computers can use for reasoning and decision-making processes. This standardized approach offers a novel method for managing network-related data and applying it to higher-level applications.

Using an ontology for network information ensures that data from diverse sources is unified and organized. To facilitate applications with specific requirements, the DKG includes a SPARQL engine, which serves as an endpoint for querying relevant information directly from the knowledge base.

\section{DKG for Real-Time Communication and Resource Management in Swarm Networks}\label{use cases}
This section explores the potential of DKG for real-time communication and resource management in swarm networks. Examples will be provided regarding the types of information that can and should be stored in DKG in section \ref{collection}, and potential use cases of DKG in swarm communication networks are discussed in section \ref{usecase}.

\subsection{Metadata Collection in RDF format}\label{collection}
\begin{figure}[ht!]
    \centering
    \includegraphics[width=0.6\textwidth]{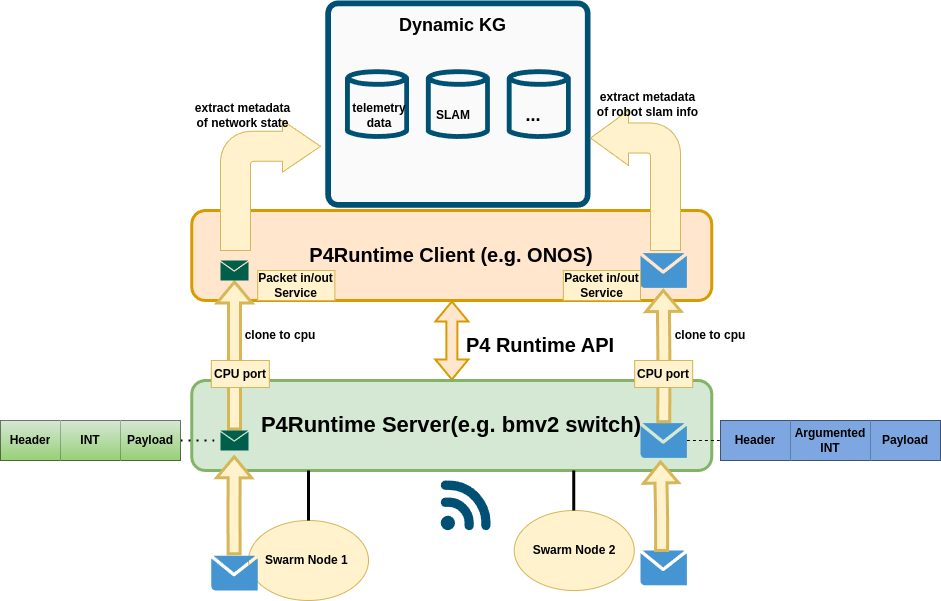}
    \caption{RDFizing Switch-specific and Swarm Node-specific Metadata}
    \label{fig:dkg-info}
\end{figure}

Figure~\ref{fig:dkg-info} illustrates the annotation of two categories of metadata in RDF format: switch-specific and swarm node-specific.
Switch-specific metadata typically includes data gathered through In-Band Network Telemetry (INT)~\cite{INTSpecs}, a robust diagnostic technique capable of measuring and recording the performance of each marked packet traversing the network. Detailed switch information, such as switch ID and hop latency, resided in the packet header and can be retrieved using P4's intrinsic functions like send\_to\_cpu or copy\_to\_cpu for header extraction. Alternatively, the DKG may offer an API for storing INT metadata collected by an INT metadata collector. The former solution involves communication through the controller, while the latter relies on direct communication between the INT collector and the DKG. 

An innovative design for INT can be activated for all swarm nodes. Besides traditional network-specific metadata such as switch ID and hop latency, this new method is aimed to embed additional augmented metadata fields related to the properties of swarm nodes such as CPU load, device localization, or capability information into the packet header. These rich metadata regarding both the network state and the swarm node are cloned to the controller via packet I/O operations for further inspection and are extracted as RDF objects inside the ONOS controller. They are then stored as RDF triples for the DKGs at the controller node. The controller node can be single or multiple; thus, the DKGs can be centralized or distributed. Taking the distributed controllers as an example, the remote controller node can maintain a partial view of the network state, based on which the distributed controllers hold distributed DKGs of the whole network, making it possible to store and query network information in a distributed fashion. In this way, remote controllers work collaboratively, such as figuring out the optimal path selection rule for devices in the data plane by sharing each one’s knowledge of the current network state. Via P4Runtime, the controller then installs the updated path selection rule, i.e., modified table entries, into the P4Runtime Server, that is the P4 Switch.

\subsection{Use Case in Swarm Communication Networks}\label{usecase}
Next, examples of use cases in swarm communication will be presented to demonstrate the functionality and benefits that DKG can provide to swarm networks in practical scenarios.
This design of the DKG is particularly beneficial for use cases involving robots or autonomous vehicles (referred to as swarm nodes) that move dynamically within the same swarm or among various swarms. Data Distribution Service (DDS) \footnote{\url{https://www.omg.org/spec/DDS/1.4/PDF}}facilitates real-time communication between these swarm nodes, utilizing the Real-Time Publish-Subscribe (RTPS) \footnote{\url{https://www.omg.org/spec/DDSI-RTPS/2.2/PDF}} wire protocol for network communication.

\begin{figure}[ht!]
    \centering
    \includegraphics[width=0.6\textwidth]{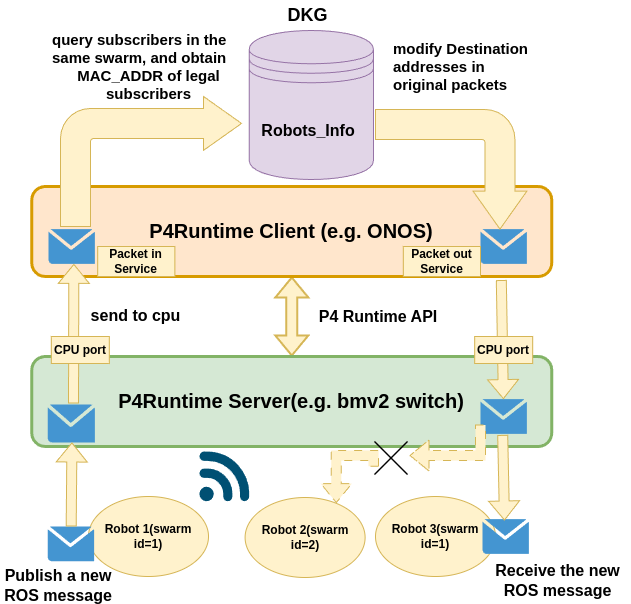}
    \caption{DKG Use Case: Access Control Using the DDS Framework}
    \label{fig:access control in ros}
\end{figure}

For instance, Figure~\ref{fig:access control in ros} depicts a scenario where a publisher sends a new RTPS message intended only for its subscribers who share the same swarm ID. As the RTPS packet passes through the P4 switch, it is forwarded to the control plane for further inspection, as required by the P4 program. The controller node queries the DKG to obtain a list of swarm nodes that belong to the same group as the publisher, identifying these as authorized subscribers. Consequently, the controller resolves the destination addresses from the knowledge graph. By leveraging the shared knowledge graph, the ROS publisher can multicast ROS messages to a specific group of ROS subscribers, instead of broadcasting messages to irrelevant receivers, thereby implementing access control for multicasting.

Additionally, the DKG, utilizing INT collection, enables upper-layer applications to adopt adaptive routing algorithms. For instance, the detailed INT data reflects the real-time network state, pinpointing exact locations of traffic congestion. Based on this information, the ONOS controller can update the table entries of the P4 switch to adjust end-to-end routes accordingly. Furthermore, by indicating the capabilities of swarm nodes in the augmented header field, the DKG is aware of the underlying devices' resources, allowing it to allocate resources or even achieve load balance through more intelligent decisions.

\section{Conclusion} \label{conclusion}

In summary, this paper emphasizes the benefits of integrating DKG into SDN, particularly for swarm networks. It highlights DKG's role within SDN, its capability to extract relevant content from swarm networks using specific APIs, and provides detailed insights into its architecture and workflows. Additionally, through use case illustrations in access control for robot communication networks, this paper demonstrates DKG's potential applicability in SDN, especially in swarm networks. Looking ahead, future work will focus on delivering an initial implementation of this platform and deploying it in practical swarm network scenarios.

\begin{acknowledgments}
    This work is supported by the German Research Foundation (DFG) under the COSMO project (grant No. 453130567), and by the European Union's Horizon WIDERA under the grant agreement No. 101079214 (AIoTwin), and RIA research and innovation programme under the grant agreement No. 101092908 (SmartEdge). 
\end{acknowledgments}

\bibliography{main}

\appendix

\section{Additional Figures}

\begin{figure}[ht!]
    \centering
    \includegraphics[width=0.9\textwidth]{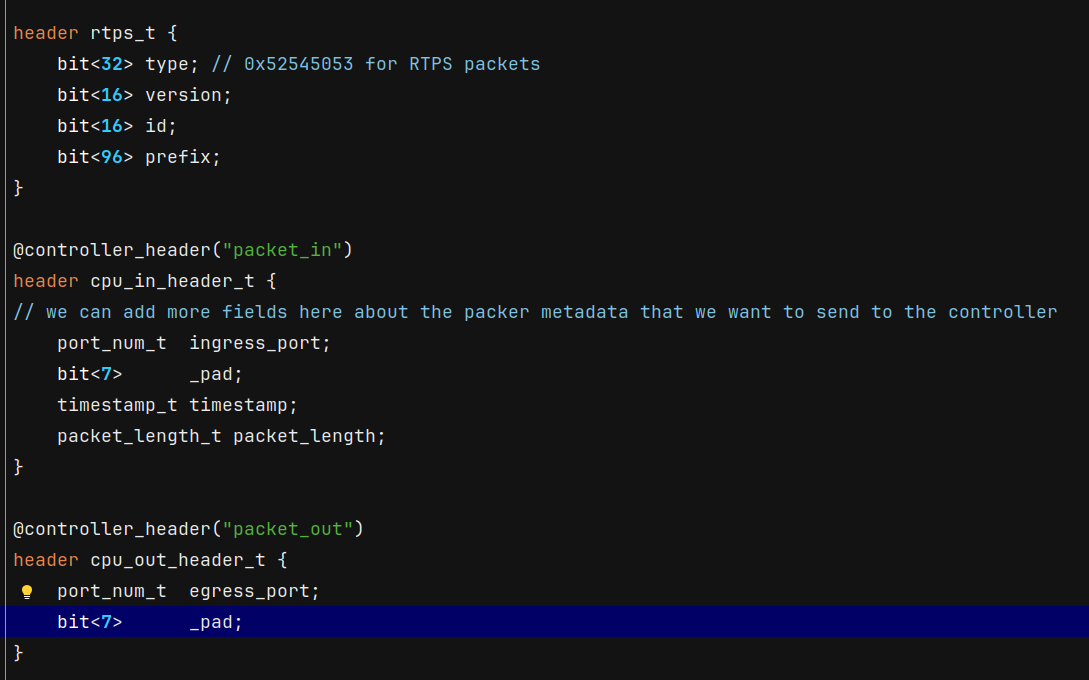}
    \caption{P4 code snippet: RTPS header, cpu\_in and cpu\_out header definition}
    \label{fig:p4 header}
\end{figure}

\begin{figure}[ht!]
    \centering
    \includegraphics[width=0.9\textwidth]{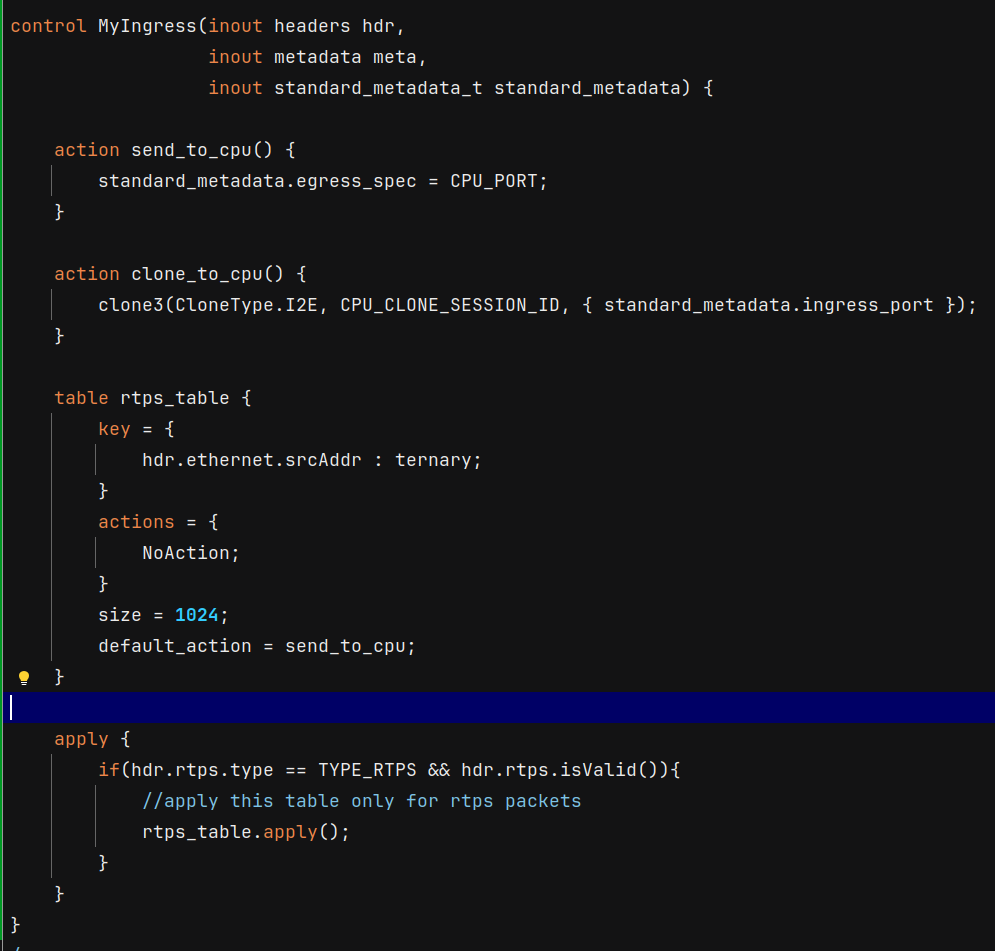}
    \caption{P4 code snippet: packet I/O service enables the RTPS packets to be sent to cpu.}
    \label{fig:p4 code}
\end{figure}



\end{document}